\PassOptionsToPackage{unicode}{hyperref}
\PassOptionsToPackage{hyphens}{url}
\documentclass[
]{article}
\usepackage{xcolor}
\usepackage[margin=1in]{geometry}
\usepackage{amsmath,amssymb}
\usepackage{iftex}
\ifPDFTeX
  \usepackage[T1]{fontenc}
  \usepackage[utf8]{inputenc}
  \usepackage{textcomp} 
\else 
  \usepackage{unicode-math} 
  \defaultfontfeatures{Scale=MatchLowercase}
  \defaultfontfeatures[\rmfamily]{Ligatures=TeX,Scale=1}
\fi
\usepackage{lmodern}
\ifPDFTeX\else
\fi
\IfFileExists{upquote.sty}{\usepackage{upquote}}{}
\IfFileExists{microtype.sty}{
  \usepackage[]{microtype}
  \UseMicrotypeSet[protrusion]{basicmath} 
}{}
\makeatletter
\@ifundefined{KOMAClassName}{
  \IfFileExists{parskip.sty}{%
    \usepackage{parskip}
  }{
    \setlength{\parindent}{0pt}
    \setlength{\parskip}{6pt plus 2pt minus 1pt}}
}{
  \KOMAoptions{parskip=half}}
\makeatother
\usepackage{graphicx}
\makeatletter
\newsavebox\pandoc@box
\newcommand*\pandocbounded[1]{
  \sbox\pandoc@box{#1}%
  \Gscale@div\@tempa{\textheight}{\dimexpr\ht\pandoc@box+\dp\pandoc@box\relax}%
  \Gscale@div\@tempb{\linewidth}{\wd\pandoc@box}%
  \ifdim\@tempb\p@<\@tempa\p@\let\@tempa\@tempb\fi
  \ifdim\@tempa\p@<\p@\scalebox{\@tempa}{\usebox\pandoc@box}%
  \else\usebox{\pandoc@box}%
  \fi%
}
\def\fps@figure{htbp}
\makeatother
\NewDocumentCommand\citeproctext{}{}

\makeatletter
 \let\@cite@ofmt\@firstofone
 \def\@biblabel#1{}
 \def\@cite#1#2{{#1\if@tempswa , #2\fi}}
\makeatother
\newlength{\cslhangindent}
\newlength{\csllabelwidth}
\newenvironment{CSLReferences}[2] 
 {\begin{list}{}{%
  \setlength{\itemindent}{0pt}
  \setlength{\leftmargin}{0pt}
  \setlength{\parsep}{0pt}
  \ifodd #1
   \setlength{\leftmargin}{\cslhangindent}
   \setlength{\itemindent}{-1\cslhangindent}
  \fi
  \setlength{\itemsep}{#2\baselineskip}}}
 {\end{list}}
\usepackage{calc}

\providecommand{\tightlist}{%
  \setlength{\itemsep}{0pt}\setlength{\parskip}{0pt}}
\usepackage{bookmark}
\IfFileExists{xurl.sty}{\usepackage{xurl}}{} 
\hypersetup{
  pdftitle={MicroTrace: A Lightweight R Tool for SNP-Based Pathogen Clustering in Outbreak Detection},
  pdfauthor={Kaitao Lai1},
  hidelinks,
  pdfcreator={LaTeX via pandoc}}

\title{MicroTrace: A Lightweight R Tool for SNP-Based Pathogen
Clustering in Outbreak Detection}
\author{Kaitao Lai\textsuperscript{1}}
\date{2025-07-02}

\begin{document}
\maketitle

\textsuperscript{1} University of Sydney

\section{Summary}\label{summary}

\textbf{MicroTrace} is an open-source R tool that performs SNP-based
hierarchical clustering to detect potential transmission clusters from
pathogen whole-genome sequencing (WGS) data. Designed for
epidemiologists, microbiologists, and genomic surveillance teams, it
processes SNP distance matrices and outputs dendrograms and cluster
tables with optional metadata integration. MicroTrace enables
reproducible outbreak detection workflows with minimal setup.

\section{Statement of Need}\label{statement-of-need}

This tool was motivated by my own experience supporting hospital
outbreak investigations where rapid, reproducible, and interpretable
clustering was required---but often slowed by the complexity or
inflexibility of available pipelines. During exploratory analyses of SNP
distance matrices, I observed that closely related isolates often
exhibited SNP distances in the lowest decile of all pairwise
comparisons. As such, MicroTrace uses the \textbf{10th percentile of SNP
distances} as a conservative default threshold for outbreak
definition---prioritizing sensitivity in early cluster detection. This
approach reflects empirical patterns in genomic epidemiology (Payne et
al. 2021) while remaining adjustable for local context.

Although whole-genome sequencing (WGS) has revolutionized pathogen
surveillance (K"oser et al. 2012), outbreak detection workflows often
remain fragmented and require specialized tools or pipelines. Tools like
Snippy (Seemann 2015) provide variant calling, and visualization
platforms such as GrapeTree offer phylogenetic context, but a
lightweight, scriptable R solution for direct clustering from SNP
matrices is lacking.

\textbf{MicroTrace} addresses this need by providing: - Automated
hierarchical clustering with configurable SNP thresholds - Distance
distribution-based threshold recommendation - Metadata integration for
spatiotemporal context - Scripted reproducibility and a markdown-based
HTML report

This simplicity supports real-time response in clinical microbiology and
public health settings.

\section{MicroTrace Design and
Features}\label{microtrace-design-and-features}

MicroTrace consists of modular R functions and follows a clear
processing workflow:

\begin{itemize}
\tightlist
\item
  \texttt{distance\_loader()}: Loads a pairwise SNP distance matrix from
  CSV
\item
  \texttt{auto\_threshold\_suggestion()}: Suggests a clustering
  threshold (e.g., 10th percentile SNP distance) and generates histogram
  and density plots
\item
  \texttt{run\_microtrace\_clustering()}: Performs UPGMA clustering,
  cuts the tree at the threshold, joins optional metadata, and generates
  output files
\end{itemize}

Additional features include: - Optional metadata integration (collection
date, ward, patient ID) - Publication-ready PNG output of dendrogram and
SNP distance plots - Markdown-based HTML report template - Intra-cluster
SNP summary statistics (mean, SD, range) - Unit tests for all core
functions

\section{Simulated Dataset Example}\label{simulated-dataset-example}

A simulated dataset (10 samples) demonstrates two genetically distinct
clusters. The SNP distances among samples from the same hospital ward
(Ward A) are 0--3, while those from Ward B are 7--10. When clustered
with a 5-SNP threshold, MicroTrace correctly identifies two distinct
groups.

Metadata are merged into the output table, allowing alignment of genetic
clusters with sample origin. The example illustrates utility for
localized outbreak tracking in hospitals or communities.

\section{Reproducibility and Testing}\label{reproducibility-and-testing}

MicroTrace includes a \texttt{tests/} directory with automated tests
using \texttt{testthat}, covering: - Distance matrix loading - Threshold
estimation - Cluster output generation - Output dimensions and object
class checks

This supports reproducibility and continuous integration.

\section{Visualization and Reporting}\label{visualization-and-reporting}

The \texttt{MicroTrace\_Report.Rmd} template provides a user-friendly,
customizable HTML report. It summarizes the clustering results,
visualizes the SNP distance histogram, density plot, dendrogram, and
presents metadata-aware statistics. This facilitates communication with
infection control or public health teams.

\section{Software Repository}\label{software-repository}

The source code for MicroTrace is freely available on GitHub at:\\
\url{https://github.com/biosciences/MicroTrace}

\section{Acknowledgements}\label{acknowledgements}

The idea for MicroTrace originated during my involvement in
antimicrobial resistance surveillance projects, where I observed a
recurring gap between high-throughput WGS data and pragmatic,
interpretable clustering tools usable by clinical microbiologists. This
led me to develop a self-contained R solution focused on outbreak
clustering logic, metadata overlay, and report generation. I thank the
research community of University of Sydney for fostering applied
genomics thinking and encouraging a culture of lightweight, real-time
analytics in infectious disease genomics.

The author acknowledges colleagues from University of Sydney for
insights on real-time SNP clustering in public health genomics.

\section*{References}\label{references}
\addcontentsline{toc}{section}{References}

\protect\phantomsection\label{refs}


\begin{CSLReferences}{1}{0}
\bibitem[\citeproctext]{ref-koser2012wgs}
K"oser, C. U., M. J. Ellington, E. J. Cartwright, S. H. Gillespie, N. M.
Brown, M. Farrington, M. T. G. Holden, et al. 2012. {``Routine Use of
Microbial Whole Genome Sequencing in Diagnostic and Public Health
Microbiology.''} \emph{PLoS Pathogens} 8 (8): e1002824.
\url{https://doi.org/10.1371/journal.ppat.1002824}.

\bibitem[\citeproctext]{ref-payne2021dynamic}
Payne, M., D. J. Ingle, M. Valcanis, T. Seemann, et al. 2021.
{``Enhancing Genomics-Based Outbreak Detection of Endemic Salmonella
Enterica Serovar Typhimurium Using Dynamic Thresholds.''}
\emph{Microbial Genomics} 7 (6): 000310.
\url{https://doi.org/10.1099/mgen.0.000310}.

\bibitem[\citeproctext]{ref-seemann2015snippy}
Seemann, Torsten. 2015. {``Snippy: Rapid Haploid Variant Calling and
Core Genome Alignment.''} \url{https://github.com/tseemann/snippy}.

\end{CSLReferences}
\end{document}